# Atmospheric Aerosol Limb Scanning
# Based on the Lunar Eclipses Photometry

Oleg S. Ugolnikov [a,b,*] and Igor A. Maslov [a,c]

[a] *Space Research Institute, Profsoyuznaya st., 84/32, 117997, Moscow, Russia*
[b] *Astro-Space Center, Lebedev's Physical Institute, Profsoyuznaya st., 84/32, 117997, Moscow, Russia*
[c] *Sternberg Astronomical Institute, Universitetsky prosp., 13, 119992, Moscow, Russia*

E-mail: *ugol@tanatos.asc.rssi.ru*

**Abstract.** The work is devoted to the analysis of the surface photometric observations of two total lunar eclipses in 2004. The lunar surface relative brightness distribution inside the umbra was used to retrieve the vertical distribution of aerosol extinction of the solar radiation expanding by a tangent path and its dependence on the location at the limb of the Earth. The upper altitude of troposphere aerosol layer was estimated for different latitude zones. The correlation between additional aerosol extinction in the upper troposphere and cyclones was investigated.

## 1. Introduction

The problem of optical investigations of the Earth's atmosphere at different altitudes is quite actual. The changes in ozone layer, trace gases and aerosol content make necessary to carry out the measurements of atmosphere condition in different locations on the Earth and different atmosphere layers. Optical sounding – photometry, polarimetry, spectroscopy – is one of most effective ways to solve this problem. Usually the atmosphere parameters being measured from the ground is the integration by the definite range of altitudes, basically by dense and optically thick near-ground layers. In order to research the upper atmosphere layers separately, the lidar or twilight sounding can be used.

One more effective method to investigate the optical properties of medium and high atmosphere layers is to measure the light absorption, emission and scattering by the tangent ray path with different tangent point altitudes. This case the most part of the path has an almost the same altitude, the influence of higher layers with less density is sufficiently smaller than in the case of zenith measurements from the ground. Such path has larger optical depth, this gives us the possibility to detect small optical features better than by other methods.

This advantage is principal for the investigations of trace gases in the atmosphere those had been made in a number of space programs, and the SCIAMACHY experiment [1-3] is worthy of note. Alongside with nadir measurements of underlying atmosphere, it holds two kinds of tangent ray investigations: the spectral measurement of scattered emission during the daytime and absorption of the Sun or Moon radiation at their rise at the limb of the Earth. This makes possible to retrieve the distribution of a large number of gases in the atmosphere, clouds and aerosol.

However, the tangent radiation measurements are possible on the ground if we observe the body moving through the shadow of the Earth. This body can be the artificial satellite of the Earth, this case we will measure one extinction profile above the definite location on the planet. But if the shadow is crossed by the Moon, the differential surface photometry will give the possibility to build the vertical profile of extinction coefficient along the part of the Earth's limb. The size of this part will depend on the path of the Moon through the shadow (or umbra). The optical theory of lunar eclipse was developed in the middle of XX century, wide review was performed by Link [4]. In our work we will develop the method of aerosol extinction calculation and apply it to the results of observation of two lunar eclipses.

* Corresponding author. Fax: +7-095-333-5178.

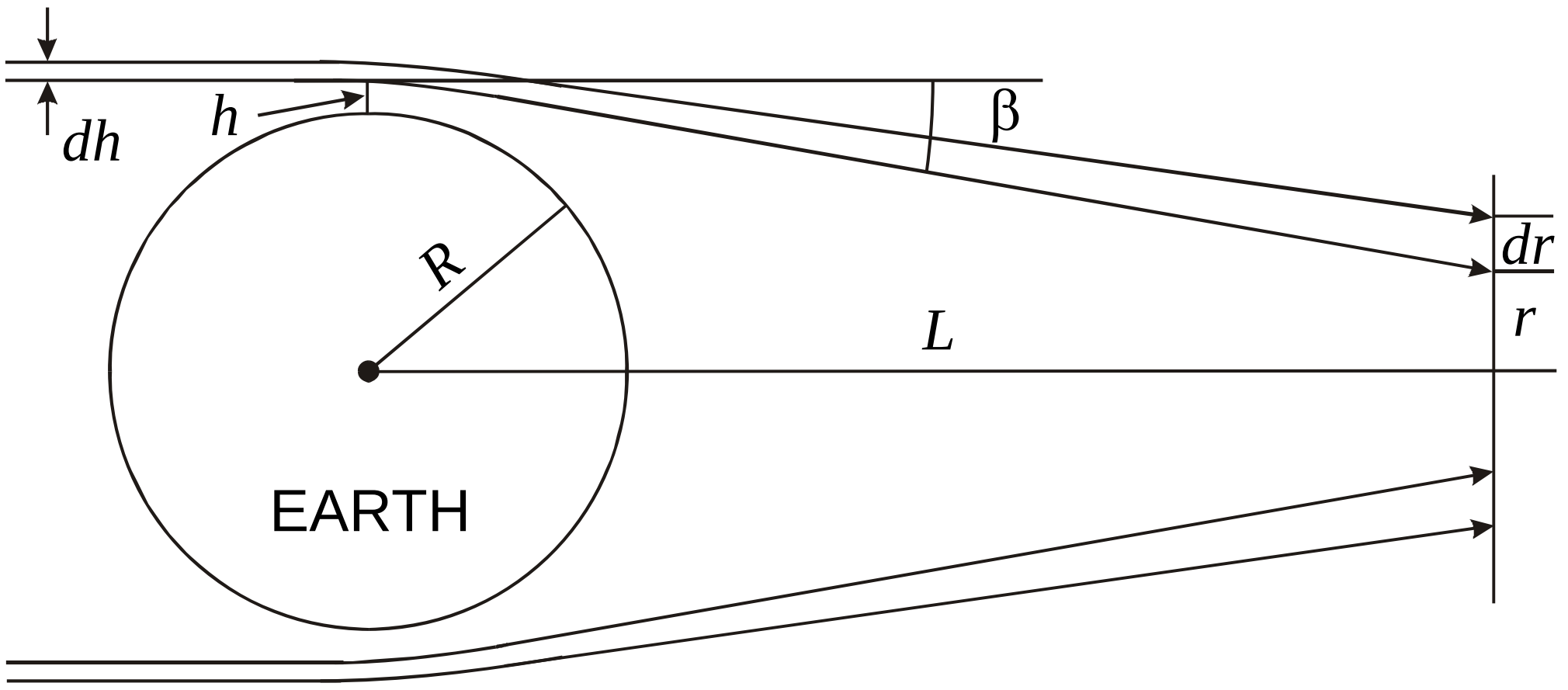

Figure 1. Geometry of radiation transfer in the shadow of the Earth.

**2. Radiation transfer during the lunar eclipse**

Describing the radiation transfer, we will use the same theory as in [4] with two acceptable simplifications: we assume that the parallax of the Sun is extremely less than the one of the Moon and the troposphere of the Earth is thin compared with the size of our planet.

Let's imagine that the radiation from the distant point-like object with energy flux $F$ is transferring horizontally above the Earth's surface (see Figure 1). Since the density and refracting index of atmosphere substance is decreasing with the altitude, the ray changes its direction turning to the geometric shadow area. Let's assume that there is no extinction and scattering in the atmosphere and calculate the energy flux inside the umbra at the distance $L$ from the center of the Earth. The amount of energy going through the ring around the Earth with radius ($R$+$h$) and thickness $dh$ during one second is equal to

$$dE = 2\pi F\,(R+h)\,dh. \tag{1}$$

We assume that the refraction angle is small and the radiation is transferred almost horizontally in the atmosphere. This case the ray is refracting by the angle $\beta(h)$ that can be calculated as

$$\beta(h) = -\int \frac{dn(h_S)}{dh_S}\,ds, \tag{2}$$

where $n$ is the refraction index of atmospheric substance, $ds$ is the element of the ray path and $h_S$ is the altitude of this element. The integration is making by the whole ray path in the atmosphere. This formula gives the possibility to calculate the path and the value of $\beta$ simultaneously. Passing by the distance $L$ to the screen plane, the radiation will fill the ring with radius

$$r = \left| R + h - L\beta(h) \right| \tag{3}$$

and thickness

$$dr = dh\left| 1 - L\frac{d\beta}{dh} \right|. \tag{4}$$

In the atmosphere conditions, where density is decreasing with the altitude, the derivative $d\beta/dh$ is always negative and we may drop the module sign in the last formula. It leads us to the expression of energy flux in the screen plane:

$$f = \frac{dE}{2\pi r dr} = F \cdot \left| \frac{\pi_0}{\pi_0 - \beta} \right| \cdot \frac{1}{1 - L\dfrac{d\beta}{dh}}. \qquad (5)$$

Here we assume that the atmosphere is thin compared with the Earth's size and the quantity $h$ is many times less than $R$. $\pi_0 = R/L$ means the equatorial parallax of the screen. We can see that there are two effects changing the energy flux. The first one is greater than unity and describes the "focusing" effect that may reach infinity in the center of the shadow, if the light source is point-like. The second one is less than unity describing the "refraction divergence" effect. It is important to note that in the case of distant light source the module of $(\pi_0 - \beta)$ is equal to angular distance between the light source and center of the Earth visible from the current point inside the umbra.

In the real atmosphere the energy flux gets weaker also due to the scattering and extinction in the atmosphere. This case the value of $f$ is equal to

$$f = F \cdot \left| \frac{\pi_0}{\pi_0 - \beta} \right| \cdot \frac{1}{1 - L\dfrac{d\beta}{dh}} \cdot T_G(h) \cdot T_A(h), \qquad (6)$$

where $T_G(h)$ and $T_A(h)$ is the atmosphere gas and aerosol transparency by the tangent path of the ray. As against the refraction, which is almost constant inside the observational spectral band (see the next chapter), the extinction properties strongly depends on the wavelength and the gas transparency is calculating by the integration by the observational band $P(\lambda)$:

$$T_G(h) = \frac{\int e^{-\tau_G(h,\lambda)} P(\lambda)\, d\lambda}{\int P(\lambda)\, d\lambda}. \qquad (7)$$

The optical depth $\tau_G$ is mostly contributed by Rayleigh scattering component, which can be calculated, but in several spectral regions (mostly avoided at the observations) the gaseous absorption can be also added.

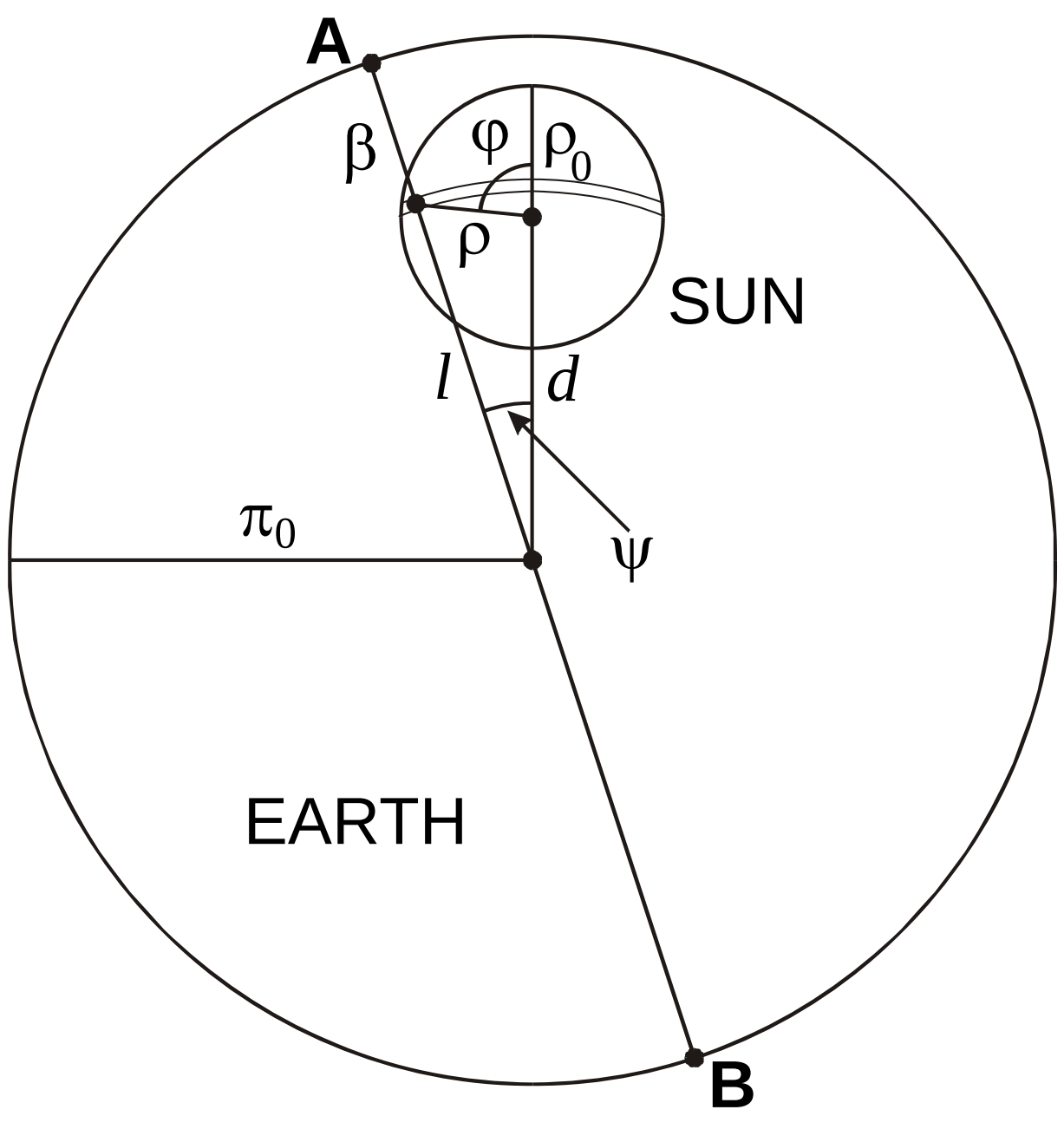

Figure 2. The eclipse picture visible from the Moon.

In real conditions the source of light is the Sun having large angular size. Let $\rho_0$ is the angular radius of the Sun. The total brightness of solar disk will be equal to

$$I_0 = \int_0^{\rho_0} \int_0^{2\pi} S(\rho)\rho\, d\varphi\, d\rho,$$

$$S(\rho) = S_0 (1 - a + a(1 - \frac{\rho}{\rho_0})^{1/2}). \qquad (8)$$

Here $S_0$ is surface brightness of the center of the solar disk, $a$ is the coefficient of darkening towards the edge of the solar disk (we may assume $a=0.5$ for our spectral band).

When eclipse occurs, being observed from the Moon, the Sun is partially or totally occulted by the Earth (see Figure 2). Let $d$ will be the angular distance between the centers of Sun and Earth's disks (and the angular distance between lunar surface point and umbra center being observed from Earth). The brightness of eclipsed Sun disk will be equal to

$$I(d) = \int_0^{\rho_0} \int_0^{2\pi} S(\rho) A(l) \rho\, d\varphi\, d\rho,$$

$$l = (d^2 + \rho^2 + 2d\rho \cos\varphi)^{1/2}. \qquad (9)$$

Here $A(l)$ is the differential umbra darkening factor, characterizing the darkening of the Sun disk element situated at the angular distance $l$ from the Earth disk center. To be registered by the observer on the Moon, the light ray must be refracted in the atmosphere near points A or B. The refraction angles $\beta_{1,2}$ for these two cases are equal to ($\pi_0 - l$) and ($\pi_0 + l$), respectively. Using the assumption of thin atmosphere and relations above, we can write the expression for differential darkening factor:

$$A(l) = 1, \quad l > \pi_0,$$

$$A(l) = \frac{\pi_0}{l} (E(\pi_0 - l) + E(\pi_0 + l)), \quad l \le \pi_0, \qquad (10)$$

where $E(\beta)$ is the ray dilution factor:

$$E(\beta) = 0, \quad \beta > \beta_0,$$

$$E(\beta) = T_G(h(\beta)) \cdot T_A(h(\beta)) \cdot \frac{1}{1 - L\frac{d\beta}{dh}}, \quad 0 \le \beta \le \beta_0,$$

$$E(\beta) = 1, \quad \beta < 0. \qquad (11)$$

Here $h(\beta)$ is the altitude of the tangent point of the ray refracting in the atmosphere by the angle $\beta$ – the function reverse to $\beta(h)$. $\beta_0$ is the refraction of the Earth-grazing ray. Last case ($\beta<0$) formally corresponds to the free ray propagation without refraction and extinction in the atmosphere (that does not happen during the total eclipse when $d<\pi_0-\rho_0$ and $l<\pi_0$ for the whole solar disk).

Dividing $I(d)$ on $I_0$, we obtain the umbra darkening factor $U(d)$, the quantity that is being measured during the observations.

The functions $T_G(h)$ and $\beta(h)$ are being calculated using the model of the molecular atmosphere with pressure at the sea level equal to 1 atm and troposphere temperature distribution shown by the solid line in the Figure 3. We can use such simple model because the calculations show that the theoretical values of $U(d)$ for gaseous atmosphere are quite insensitive for the temperature distribution, just slightly changing for the one shown by the dashed line in the same figure. The stratosphere and mesosphere condition does not sufficiently influence on the dependencies $\beta(h)$ and $U(d)$.

In a time of eclipse the Moon is playing the role of the screen. The Moon radius is many times less than the distance $L$ and we can use the formula (11) obtained for a flat screen. The surface of the Moon is not white and albedo is varying over the surface, the Moon is also not absolutely spherical. But we will compare the brightness of the region on the Moon during the eclipse only with the one in the same region out of eclipse at the fixed angular distance from the umbra center (1.6° was chosen). However, the phase angle during the eclipse can differ about 0.5° – 1° from such conditions that can lead to brightness change [5], but almost constant value of lunar brightness outside penumbra (it changes on less than 2% per 1 degree of distance from umbra center) after the eclipse of May, 4, 2004, shows the insufficiency of this effect for our work.

## 3. Observations

### *3.1. General Description*

CCD surface photometry of the Moon was conducted during the total eclipses of May, 4, and October, 28, 2004 at Crimean Laboratory of Moscow Sternberg Astronomical Institute (Crimea, Ukraine). The observations were made by the camera with "Rubinar-500" lens (focal distance is equal to 500 mm, 1:8) with ST-6 CCD detector (375x241), which gave the possibility to place entire lunar disk in the frame. The exposure time varied from 0.03 sec for non-eclipsed Moon to 10 sec in the middle of totality.

Double-band interferential filter was used during the observations; both peaks avoid strong absorption bands in the Earth's atmosphere. The weak ozone Shappuis absorption in this region was taken into account using [6]. The instrumental spectral curve built with account of spectra of solar emission, lunar albedo (based on unpublished C. Pieters catalogue and kindly provided by the authors [7]), atmosphere scattering and absorption bands is shown in the Figure 4. Just in the interval between 900 and 1000 nm the curve is affected by water vapor band. Since the spectral interval was selected to be wide (in order to increase the S/N ratio and decrease the exposure), we must integrate the value $T_G(\lambda)$ described above multiplied by the instrumental function $P(\lambda)$ in order to account the Forbes effect which can be sufficient since the tangent optical depth of the atmosphere $\tau_G$ is large.

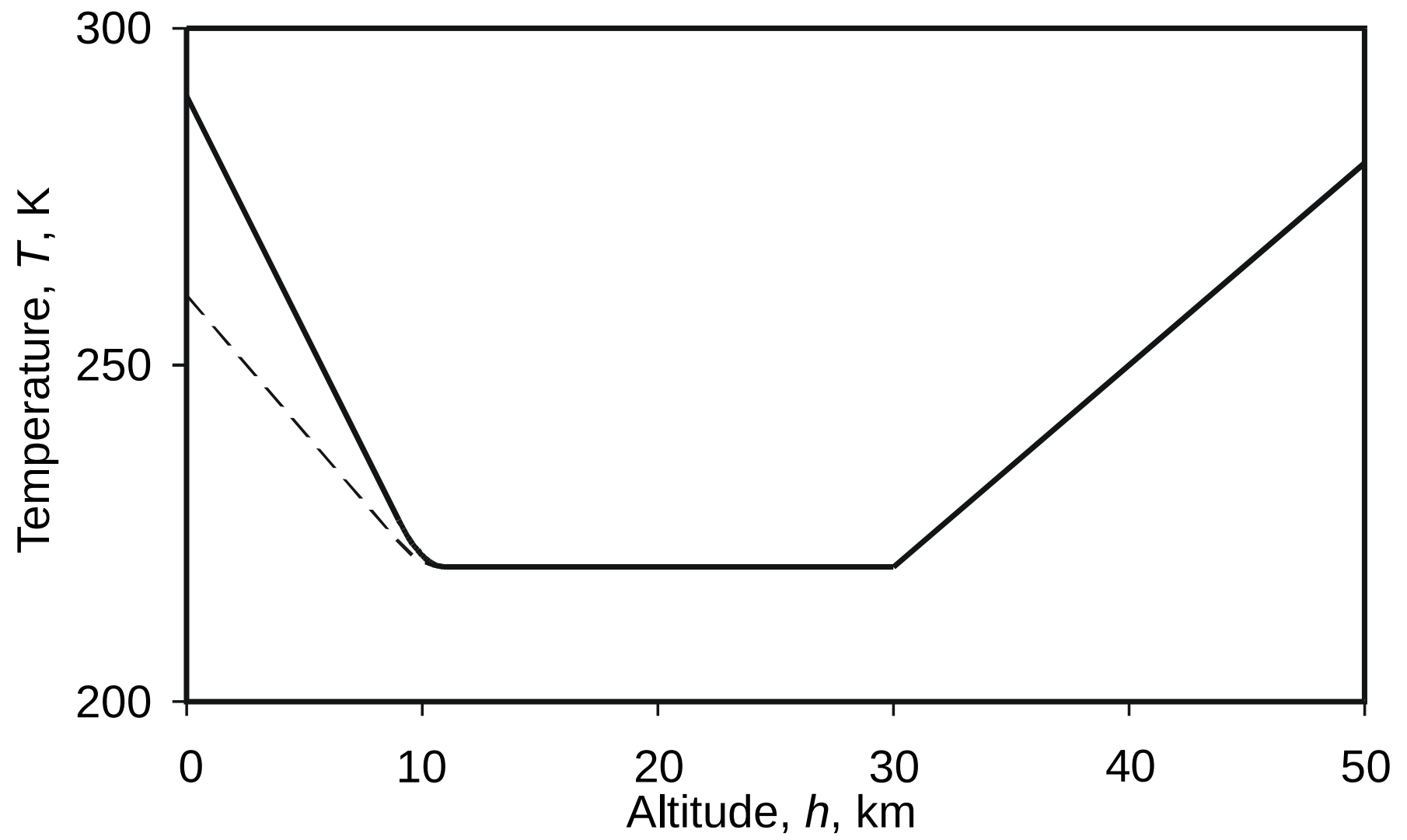

Figure 3. Model temperature vertical distribution in the atmosphere.

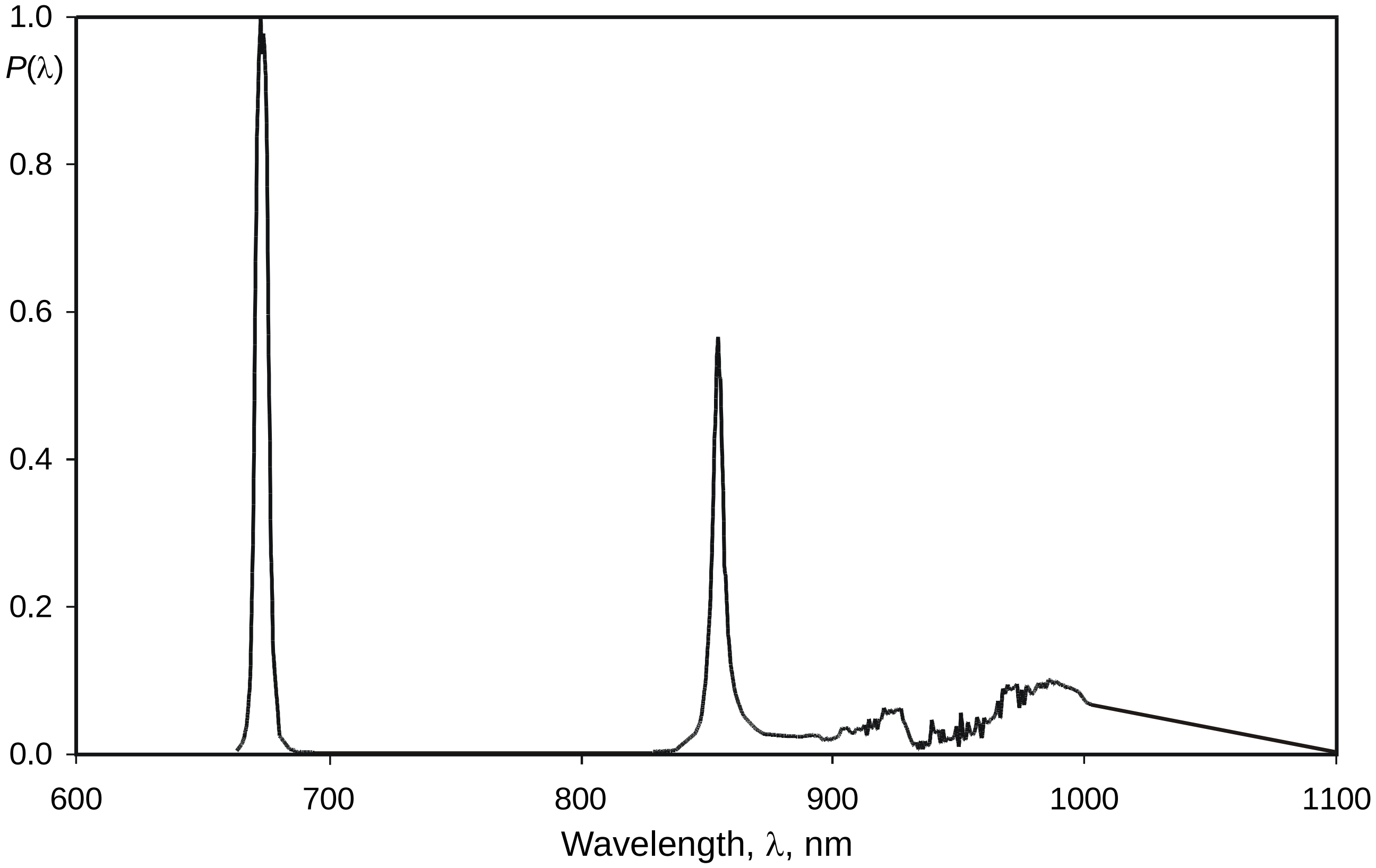

Figure 4. The spectral band of the observations with account of spectra of Sun, lunar albedo and atmosphere transparency.

Figure 5 shows the stages of both eclipses those were covered by the observations. During the eclipse on May, 04, the whole umbral area passed by the Moon was photometrically scanned. Short breaks in a sequence are the sessions of photometric standard (the star α Librae) measurements conducted for atmosphere transparency control. The eclipse of October, 28 was observed only in its starting stage with less umbra coverage and unstable weather conditions (partially cloudy sky). In spite of photometric standard measurements on the same altitude over horizon (the star γ Ceti) the results can be affected by the atmosphere transparency instability and we will consider them only for character comparison with the good quality data of May eclipse.

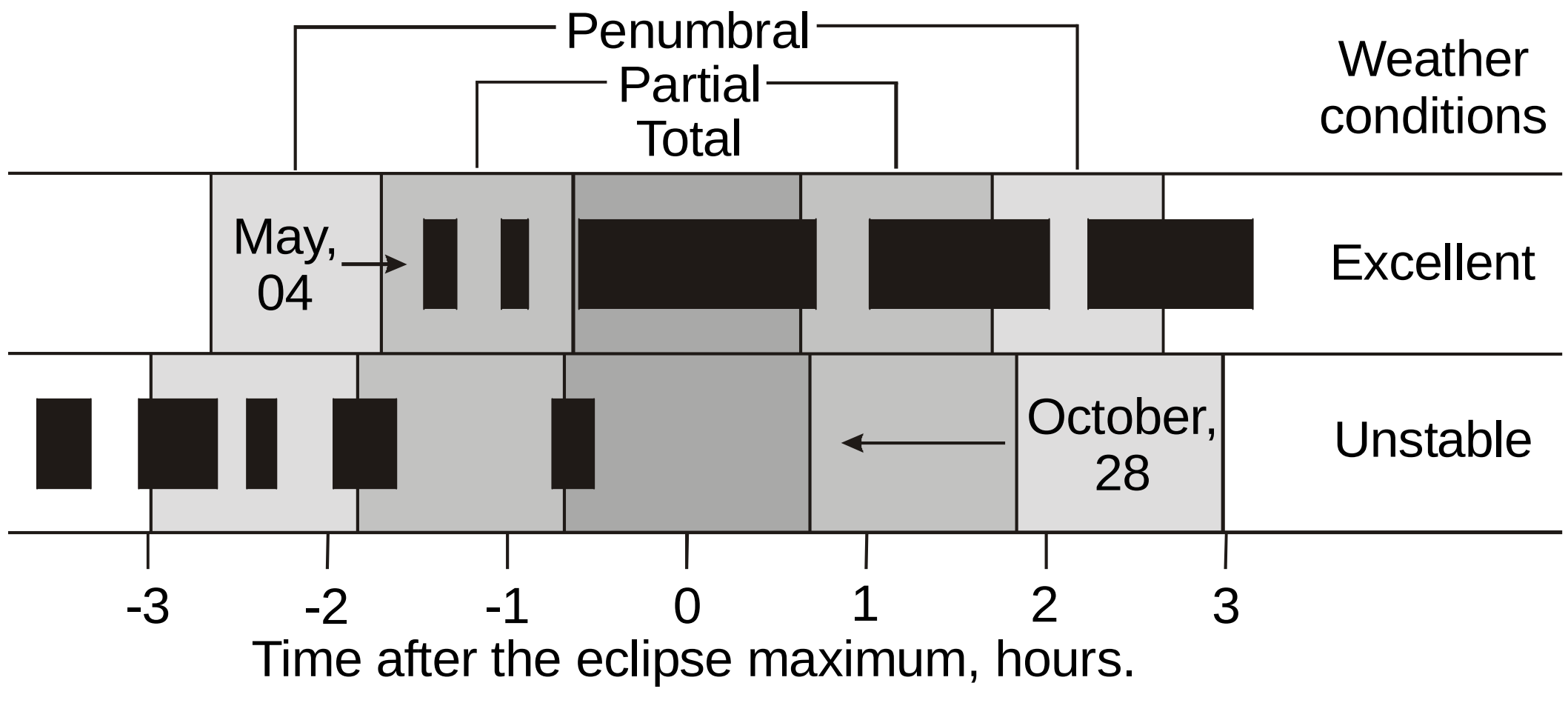

Figure 5. The Moon measurements periods (black) during both eclipses in 2004.

### 3.2. Background Subtraction

It is the basic problem we deal with analyzing the data. The influence of sky background during the lunar eclipse increases due to less measured brightness and rapid changes of the surface and scattered light intensity. For example, we cannot use the measured umbral data during the partial eclipse because of intensive and variable scattered light halo of penumbral part of lunar disk, and it is the reason of data loss at the edge of umbra and penumbra. Inside-umbra variability can also influence the scattered background, the effect we have to take into account. Since the scale of background variability is many times less than the angular size of the Moon, we cannot measure and subtract the background directly and have to build the background model.

In this model we assume that during the non-eclipse period the measured brightness of lunar surface $J_{L0}(r_L,\varphi_L)$ consists of surface brightness itself $I_{L0}(r_L,\varphi_L)$ and two background components: narrow one $B_{10}(r_L,\varphi_L)$, formed by the atmosphere scattering of nearby lunar emission converted with point-spread function of CCD-detector, and wide component $B_{20}(r_L,\varphi_L)$, which scale is comparable or larger than the size of the Moon. Here $r_L$ and $\varphi_L$ are coordinates of the point on the Moon. For lunar surface (except the near-edge areas, which are not being considered) the first component is proportional to lunar surface brightness

$$B_{10}(r_L,\varphi_L) = \alpha(r_L,\varphi_L)\cdot I_{L0}(r_L,\varphi_L), \tag{12}$$

where the ratio $\alpha(r_L,\varphi_L)$ is slowly changing across the lunar disk (and may be assumed to be constant in its small part). We may also assume $B_{20}$ to be constant in the same part. So, if we take this small sky area (0.05°x0.05° was actually taken), the measured brightness distribution inside it will be as follows:

$$J_{L0}(r_L,\varphi_L) = I_{L0}(r_L,\varphi_L)\cdot(1+\alpha) + B_{20}. \tag{13}$$

We can also note that outside the umbra (and even inside the penumbra) the second term in this formula is many times less than the first one. The situation inside the umbra gets more complicated. Due to lunar disk brightness changes from the center to the edge of the umbra, the background changes by the same direction, so the last formula will change to

$$J_L(r_L,\varphi_L,d) = I_L(r_L,\varphi_L,d)\cdot(1+\alpha) + B_{21} + B_{22}d. \tag{14}$$

Taking into account that $I_L/I_{L0}=U(d)$ and building the relation between measured surface brightness inside and outside umbra, we obtain:

$$J_L(r_L,\varphi_L,d) = U(d)\cdot J_{L0}(r_L,\varphi_L) - U(d)\cdot B_{20} + B_{21} + B_{22}d. \tag{15}$$

Since we consider the small area on the lunar surface far from the edge of umbra and penumbra, we can change $U(d)$ to $U(d_0)$, where $d_0$ is the distance between the umbra and our area centers. Taking the experimental values of $J_L$ and $J_{L0}$ pixel by pixel, we build the linear relation

$$J_L(r_L,\varphi_L,d) = K_1\cdot J_{L0}(r_L,\varphi_L) + K_2 + K_3 d, \tag{16}$$

calculating the quantities $K_1$, $K_2$ and $K_3$ by the minimum squares method. The topographic changes of lunar albedo lead to the synchronic variations of $J_L$ and $J_{L0}$, increasing the accuracy of such approximation. The quantity $K_1$ is the umbra darkening factor $U(d_0)$ that we have to find. However, the accuracy of this approximation is not always good, and we consider only the areas for which $E_{K1}/K_1<0.02$, where $E_{K1}$ is the error of quantity $K_1$ obtained by the minimum squares method.

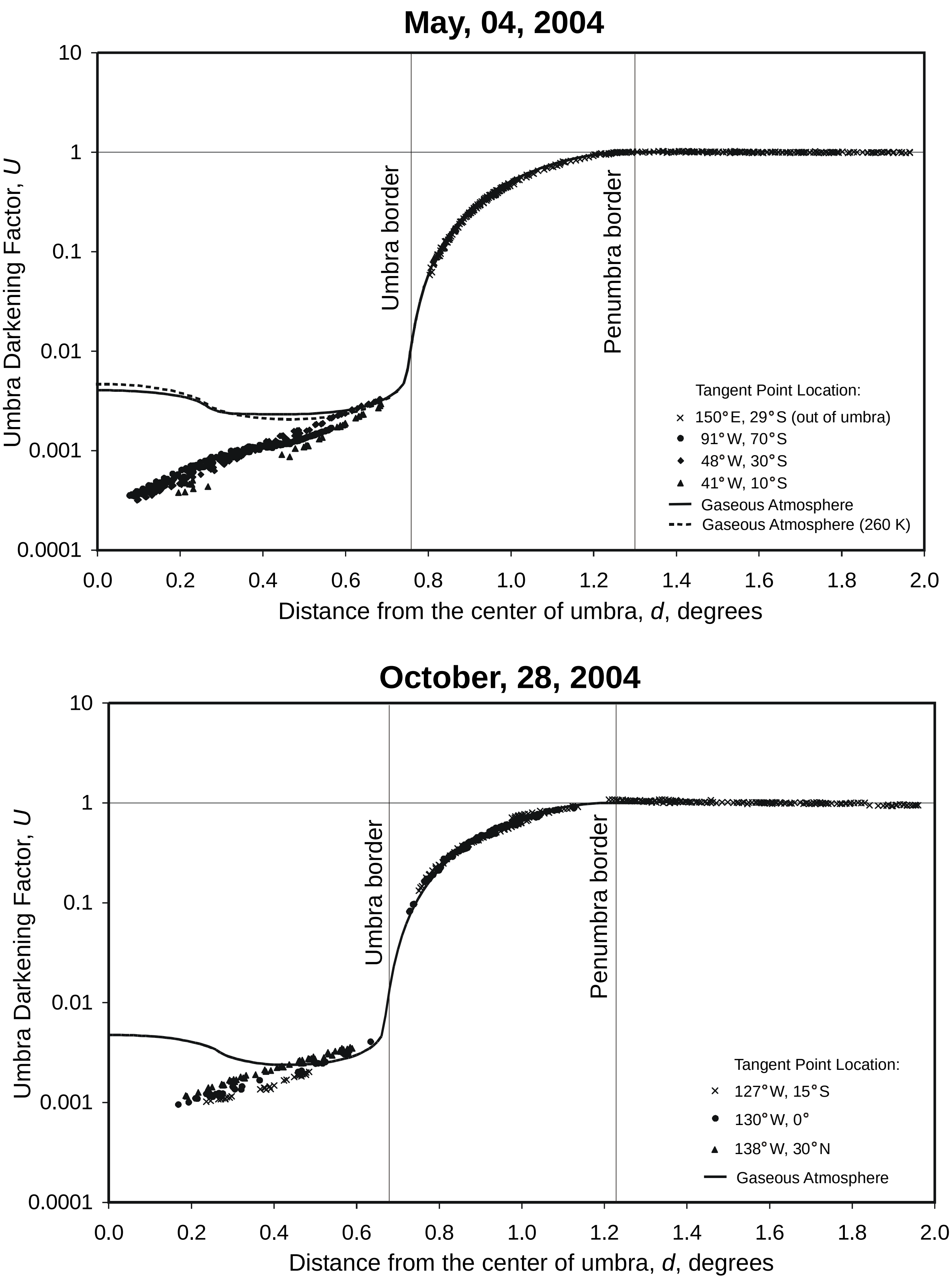

Figure 6. Umbra and penumbra profiles for both 2004 eclipses compared with numerical simulations for a gaseous atmosphere.

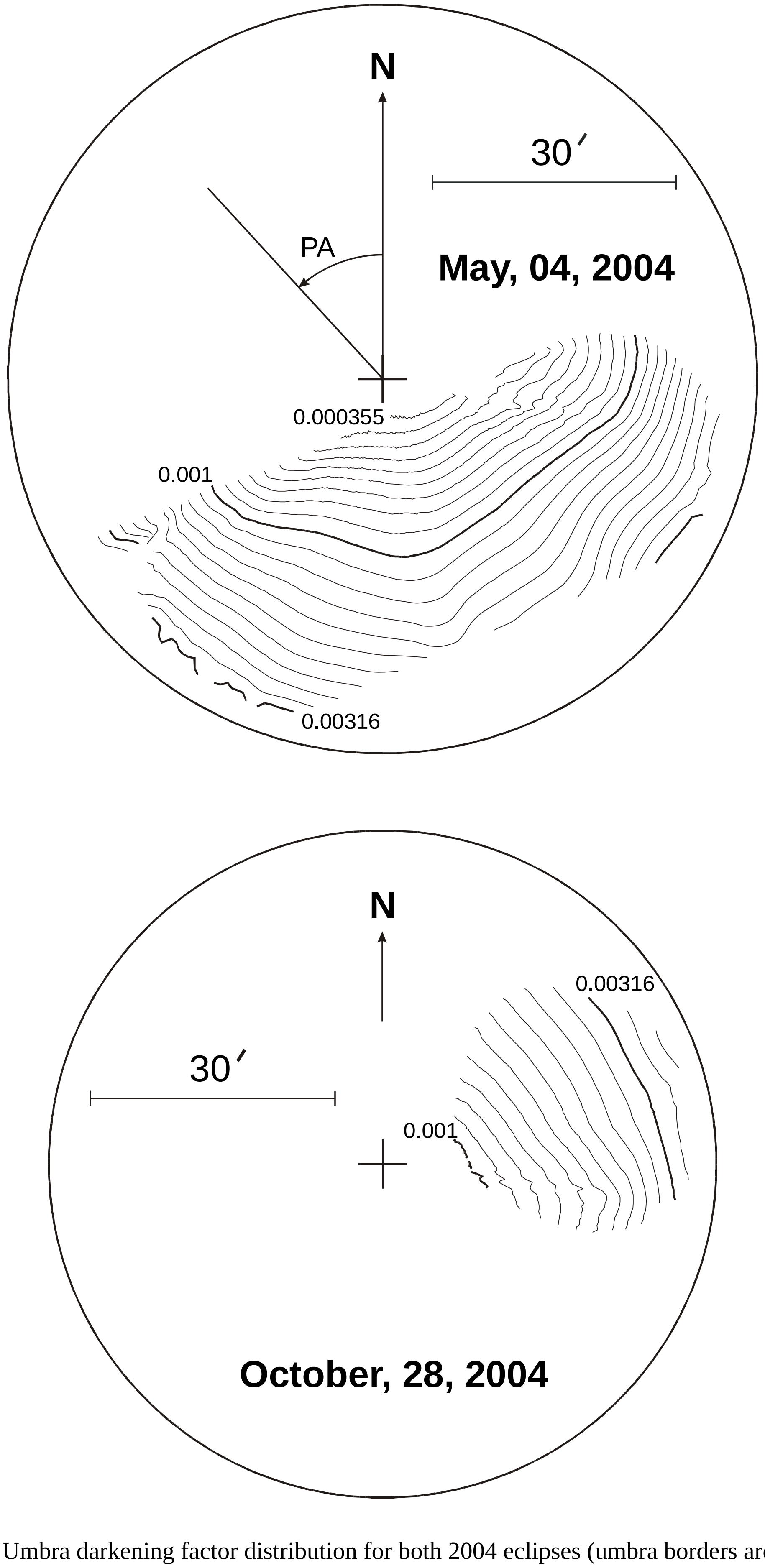

Figure 7. Umbra darkening factor distribution for both 2004 eclipses (umbra borders are shown).

*3.3. Initial Results*

Figure 6 shows the dependency of umbra darkening factor $U(d)$ on angular distance from the center of umbra compared with "gaseous" theoretical curve calculated for the case $T_A=1$ (no aerosol extinction) for both 2004 eclipses. The different types of experimental dots are corresponding to different position angles relatively umbra center or different regions on the Earth's limb where refracted radiation transfers. The coordinates of these regions are shown in the figure. The dashed line in the figure of May eclipse corresponds to the model with temperature distribution shown as dashed line in the Figure 3. We can see that the difference between two theoretical values is many times less than the difference between theoretical and observed ones. It is the reason of possibility to use the uniform temperature distribution for model calculations.

The observational and theoretical data coincide outside the eclipse and inside penumbra, where the Sun is not covered or partially covered by the Earth. It is not surprising since the atmosphere does not violate the Moon brightness during this time and this brightness is defined by geometrical picture only. Noticeable deviations for October eclipse can be explained by photometric errors due to unstable weather conditions. We can also see good agreement in the outer regions of umbra, illuminated by the solar emission refracted by small angles in the upper troposphere layers (experimental dots on October, 28 lie even higher, but it seems to be the effect of bad quality of photometry again). Thus, $T_A=1$ in the case of small refracting angle $\beta(h)$. In the inner regions of umbra experimental brightness of lunar surface is sufficiently less than theoretical values that can be explained by aerosol extinction ($T_A<1$). Theoretical curve contains noticeable "jump" at $d$ about 0.3° related with appearance of second path of solar radiation transfer to the deep umbra region (passing point B in the Figure 2). But this jump is unseen in the experimental data, so we may conclude that $T_A(h)$ vanishes near the ground ($h=0$). Umbra brightness ratio decreases to the center quite fast, so the influence of scattered light (that would make this decrease slower) is found to be insufficient.

The general distribution of umbra darkening factor inside the lunar trace in the umbra (or its observed part in October) is shown in the Figure 7. First effect that is easy to see and was noticed earlier [4] is absence of radial symmetry observed for both eclipses. The isophotes are near to ellipses extended in the equatorial direction, where umbra is darker. Local darkening near position angle 190° at May, 4, is worthy of note. This region corresponds to the tangent point location 91° W, 70° S (South-West Atlantic) with dots shown in the Figure 6.

## 4. Aerosol extinction calculation

*4.1. Mathematical Procedure*

Let us consider the case of total lunar eclipse ($d<\pi_0-\rho_0$) and express the formula (9) in terms of refraction angles $\beta_{1,2}$:

$$I(d)=\int_0^{\rho_0}\int_0^{2\pi} S(\rho)\left(\frac{\pi_0}{\pi_0-\beta_1}E(\beta_1)+\frac{\pi_0}{\beta_2-\pi_0}E(\beta_2)\right)\rho\, d\varphi\, d\rho,$$
$$\beta_{1,2}=\beta_{1,2}(d,\rho,\varphi). \tag{17}$$

The value of $\beta_1$ can change from 0 to $\pi_0$, and the value of $\beta_2$ can change from $\pi_0$ to $\beta_0$, these ranges do not overlap. Basing on the properties of delta-function, we may rewrite this formula as follows:

$$I(d)=\int_0^{\pi_0}\int_0^{\rho_0}\int_0^{2\pi}\delta(\beta-\beta_1)S(\rho)\frac{\pi_0}{\pi_0-\beta}E(\beta)\rho\,d\varphi\,d\rho\,d\beta+$$

$$+\int_{\pi_0}^{\beta_0}\int_0^{\rho_0}\int_0^{2\pi}\delta(\beta-\beta_2)S(\rho)\frac{\pi_0}{\beta-\pi_0}E(\beta)\rho\,d\varphi\,d\rho\,d\beta=$$

$$=\int_0^{\beta_0}\int_0^{\rho_0}\int_0^{2\pi}\left(\delta(\beta-\beta_1)+\delta(\beta-\beta_2)\right)S(\rho)\frac{\pi_0}{\left|\beta-\pi_0\right|}T_G(h(\beta))T_A(h(\beta))\frac{1}{1-L\dfrac{d\beta}{dh}}\rho\,d\varphi\,d\rho\,d\beta. \quad (18)$$

Let us break the range of possible values of $\beta$ into the number of small intervals with center values $\beta_j$ and constant width $\Delta\beta$ ($\beta_j=j\Delta\beta$, $j$ is varying from 0 to $N=\beta_0/\Delta\beta$). We may define the quantities

$$C_j(d)=\frac{\pi_0}{\left|\beta_j-\pi_0\right|}T_G(h(\beta_j))\frac{1}{1-L\dfrac{d\beta}{dh}(\beta_j)}\int_{\beta_j-\frac{\Delta\beta}{2}}^{\beta_j+\frac{\Delta\beta}{2}}\int_0^{\rho_0}\int_0^{2\pi}\left(\delta(\beta-\beta_1)+\delta(\beta-\beta_2)\right)S(\rho)\rho\,d\varphi\,d\rho\,d\beta, \quad (19)$$

that can be calculated using the gaseous atmosphere model. Physically $C_j(d)$ is the amount of the solar arc radiation refracted by the angle ($\beta_j\pm\Delta\beta/2$) in the region of umbra where the centers of the Sun and Earth disks are situated at the angular distance $d$ from each other (this arc is shown on the solar disk in the Figure 2) in the case of gaseous atmosphere. Note that for every amount of $j$ the contribution of at least one of two delta-functions vanishes. Assuming $\Delta\beta$ to be small, we may express the total radiation amount at this point:

$$I(d)=\sum_{j=0}^{N}C_j(d)\cdot T_A(h(\beta_j)). \quad (20)$$

Having measured the umbra darkening ratio $U(d)$ in a number of points $d_i$ with the same position angle PA, we obtain the system of linear equations:

$$U_i=U(d_i)=\frac{I(d_i)}{I_0}=\sum_{j=0}^{N}C_{ij}\cdot T_j,$$

$$C_{ij}=\frac{C_j(d_i)}{I_0},\quad T_j=T_A(h(\beta_j)). \quad (21)$$

The solution of this system should be the array of $T_j$ – the values of aerosol tangent transparency of the atmosphere for different tangent point altitudes $h$ corresponding to different refraction angles $\beta_j$. But the solution meets serious problems. First problem is that we measure the function $U(d)$ only inside the interval covered by the Moon by its motion through the umbra, which angular size is smaller than umbra radius and also depends on the position angle. If we define the values of $d_i$ with the step $\Delta d=\Delta\beta$ (0.01° in our procedure), the number of equations will be less than the number of variables. However, we can increase the number of equations by decreasing $\Delta d$, but it will not bring the positive result, since the function $U(d)$ is quite smooth (see Figure 6) and the system (21) will be near to degenerate. Small experimental errors in $U_i$ will lead to huge errors in $T_j$. Degeneracy of the system is the second problem reflecting the fact that it is transformed from the first type Fredholm's integral equation (17), which is mathematically uncorrected problem. Absence of experimental data outside the lunar path does not give us the possibility to use the standard Tikhonov's regularization algorithm [8].

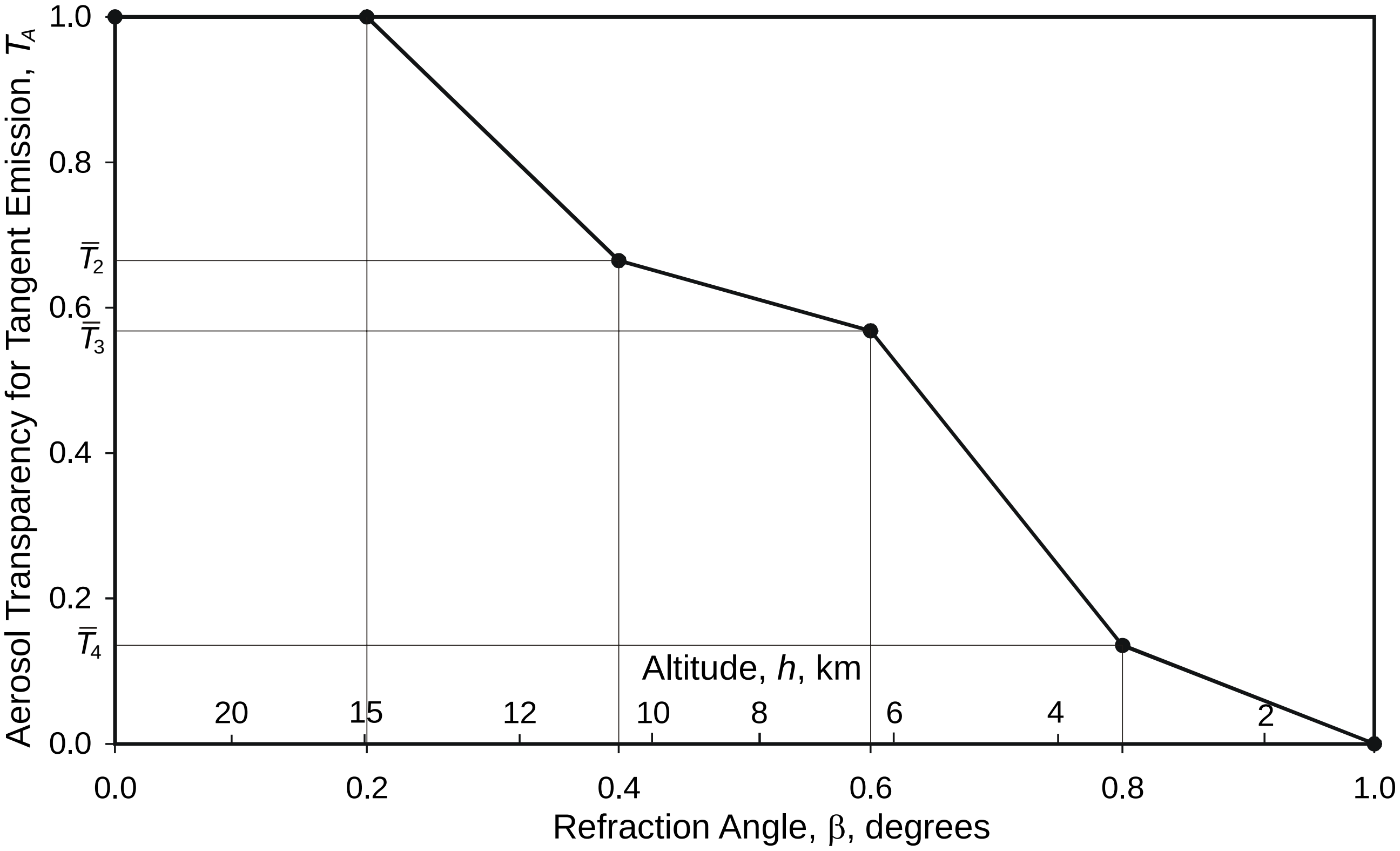

Figure 8. Function $T_A(h)$ determined by the method described in Chapter 4.1. Example shown corresponds to position angle 180° at the eclipse of May, 04.

The way to solve these problems is to decrease the number of independent parameters in the solution array $T_j$ and to take the additional assumptions concerning this array into account. First of all, we assume that the function $T(h(\beta))$ to be a broken line as shown in the Figure 8. We decrease the parameters number in $M$ times, so the array of unknown variables will take the following form:

$$\bar{T}_k = T_{kM}, \quad k = \overline{0, < N/M > +1}. \tag{22}$$

Here <$N$/$M$> means truncated integer division. Other values of $T_j$ (when $j$ is not equal to $kM$) are related with the values in the nodes of broken line in the Figure 8:

$$T_j = \bar{T}_k \frac{(k+1)M - j}{M} + \bar{T}_{k+1} \frac{j - kM}{M}, \quad k = < j/M >. \tag{23}$$

The equation system (21) will take the following form:

$$U_i = \sum_{k=0}^{<N/M>+1} \bar{C}_{ik} \cdot \bar{T}_k,$$
$$\bar{C}_{ik} = \sum_{j=(k-1)M+1}^{kM-1} \frac{C_{ij}(j-(k-1)M)}{M} + C_{ikM} + \sum_{j=kM+1}^{(k+1)M-1} \frac{C_{ij}((k+1)M - j)}{M}. \tag{24}$$

Assuming $M$=20, we decrease the resolution for the function $T_A$ by the angle $\beta$ to 0.2°. The system (24) gets soluble, but is still close to degenerate. To regularize it finally, we take into account that

$$0 \le \bar{T}_k \le 1 \tag{25}$$

for any amount of $k$, since $\bar{T}$ is the value of atmosphere transparency. As we saw above, the atmospheric aerosol near the ground is not transparent for tangent solar radiation, but upper atmosphere layers emitting outer part of contain have no absorbing aerosol (see the experimental dots in Figure 6). Taking into account that the refraction of Earth-grazing emission $\beta_0$ is equal to 1.07° for our atmosphere model, we may also assume

$$\bar{T}_6 = T(h(1.2°)) = \bar{T}_5 = T(h(1.0°)) = 0, \quad \bar{T}_0 = 1. \qquad (26)$$

After that we define variables $\bar{T}_{1-4}$ from equations (24) by minimum square method with account of (25). These values correspond to the refraction angles 0.2°, 0.4°, 0.6° and 0.8° and the tangent point altitudes 15.0, 10.6, 6.3 and 3.3 km, respectively (the altitude values depend on chosen temperature distribution, but the difference is less than 1 km). The vertical resolution being reached is thus near to the one in SCIAMACHY experiment (about 3 km [1]). The procedure is repeating for different position angles in the umbra with the step equal to 1°. For each position angle the corresponding tangent point location above the Earth is calculating, this is done for the middle of totality for May eclipse and middle of the observational period during the totality for October eclipse (see Figure 5).

*4.2. Results*

Figure 9 shows the results of the calculations described above. The graphs contain the dependencies of atmospheric aerosol tangent transparency on the position angle inside the lunar path or on latitude and longitude of the tangent point location above the Earth. The range of position angles for May eclipse is wide, the corresponding line goes through South-West Pacific, Antarctica, South-West Atlantic and eastern part of tropical and equatorial South America (see Figure 10).

The quantity of $T_A$ (15.0 km), corresponding to the refraction angle 0.2°, is turned out to be equal to unity at all position angles for both eclipses. It was expected, because the optical properties of outer umbra regions are well described by the gaseous atmosphere model, and aerosol extinction in the upper troposphere layers is very small. For brighter October eclipse the aerosol extinction is also absent for the tangent radiation refracted by 0.4° (tangent point altitude equal to 10.6 km). Here we should notice that the part of Earth limb where the refraction took place, was situated in the Eastern Pacific over the clear-sky area (little cloud cluster was near the equator as shown in the Figure 9).

Situation differs for the eclipse of May, 04. The aerosol extinction at 10.6 km appears above the large areas at the Earth's limb. The aerosol tangent transparency data are shown in the Figure 9 and in the map of clouds distribution on the Earth made in 9.5 hours after the eclipse maximum (Figure 10 [9]). We can see two large cloud clusters over the South-West Atlantic and Antarctic. In the same locations corresponding to position angles 190° and 220° the additional extinction appears (see Figure 7). We can make a conclusion that the cyclones in the southern polar latitudes expand to the upper troposphere layers. Aerosol extinction coefficient is near to $10^{-3}$ km$^{-1}$ at the altitudes about 10 km for our spectral band.

At the southern tropical latitudes the correlation of aerosol extinction and cyclones disappears, the small cloud cluster at the latitude 30°S corresponds to the maximum of quantity $T_A$. Possible explanation is the lower altitude of these clouds and the contribution of scattered light at this tangent point location. But close to equator the aerosol extinction in the upper troposphere appears again.

In the next layer being considered, with altitude 6.3 km and refraction angle 0.6°, we cannot see the correlation of $T_A$ with cloud distribution over the limb. The reason is the fact that the radiation transfers from the parts of the Sun closer to the centre the Earth while observing from the Moon. Having large angular size, the Sun covers wide range of the angles $\psi$ (see Figure 2) and parameter $T_A$ having measured is average over the large area along the limb. But anyway the sufficient increase of aerosol extinction near the equator is seen for both eclipses, not depending on the cloud distribution. Increased aerosol extinction (up to $5*10^{-3}$ km$^{-1}$) is seems to be the common property of equatorial mid-troposphere layers.

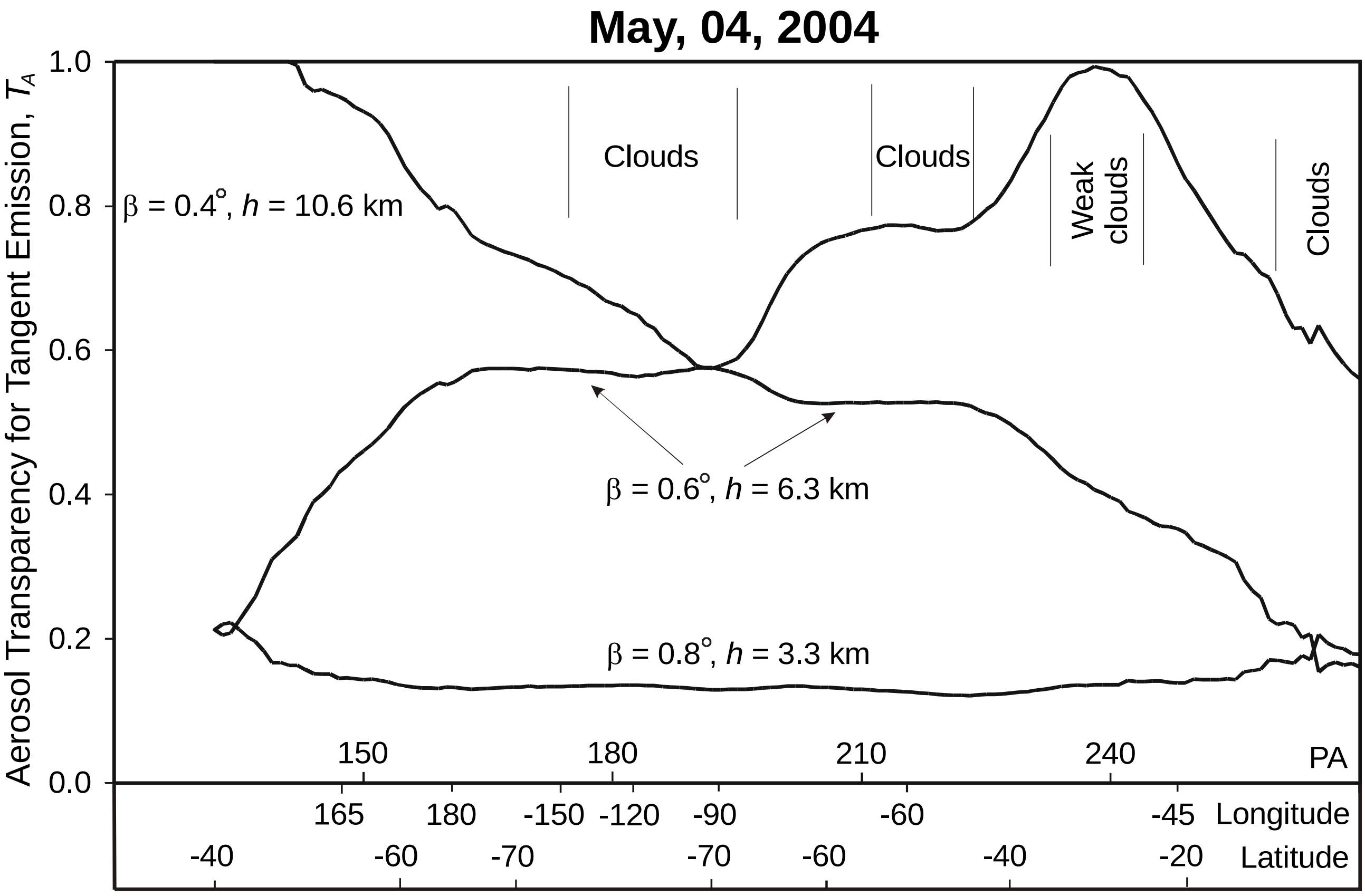

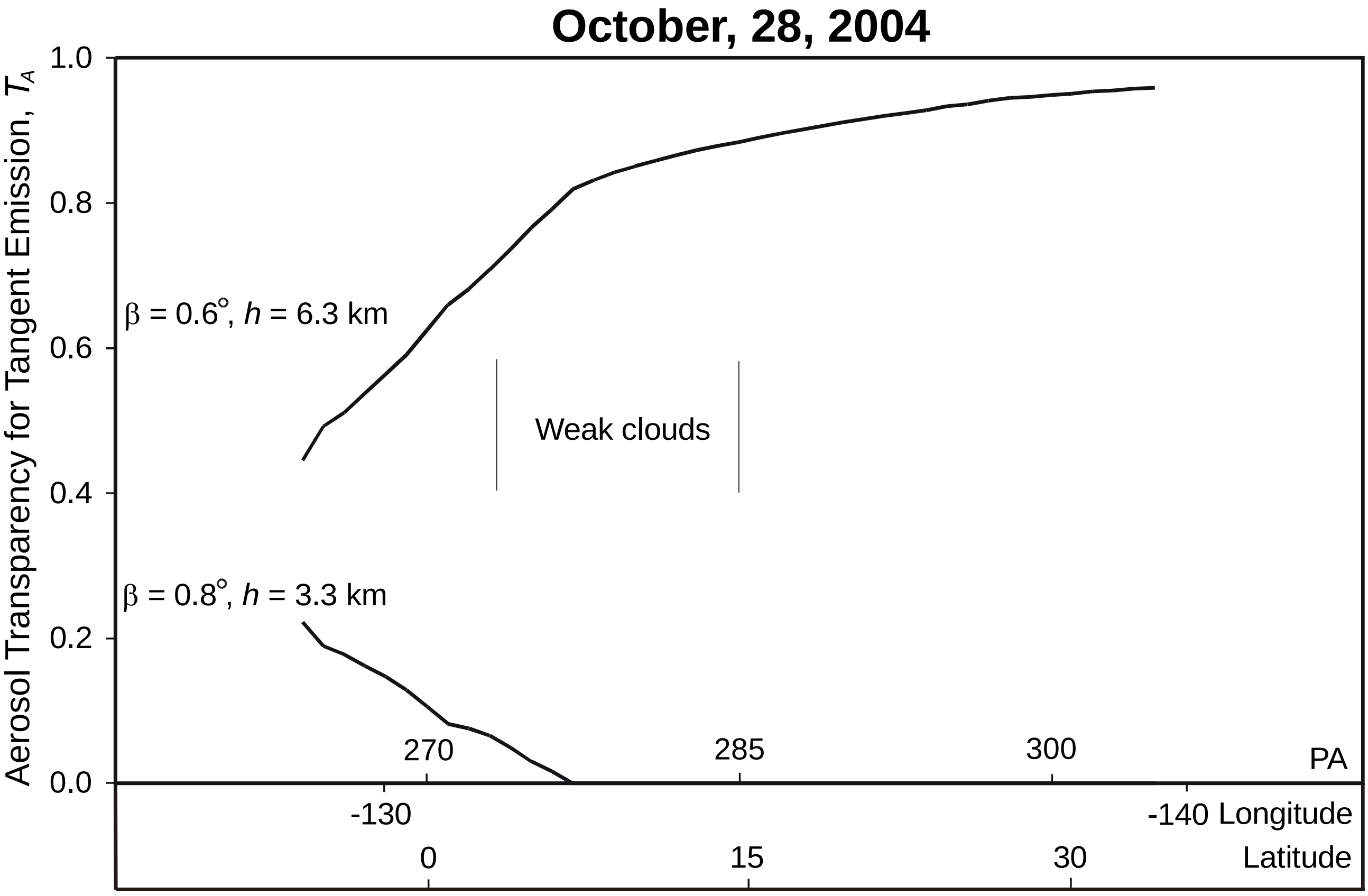

Figure 9. Atmospheric aerosol transparency for the radiation paths with different tangent altitudes.

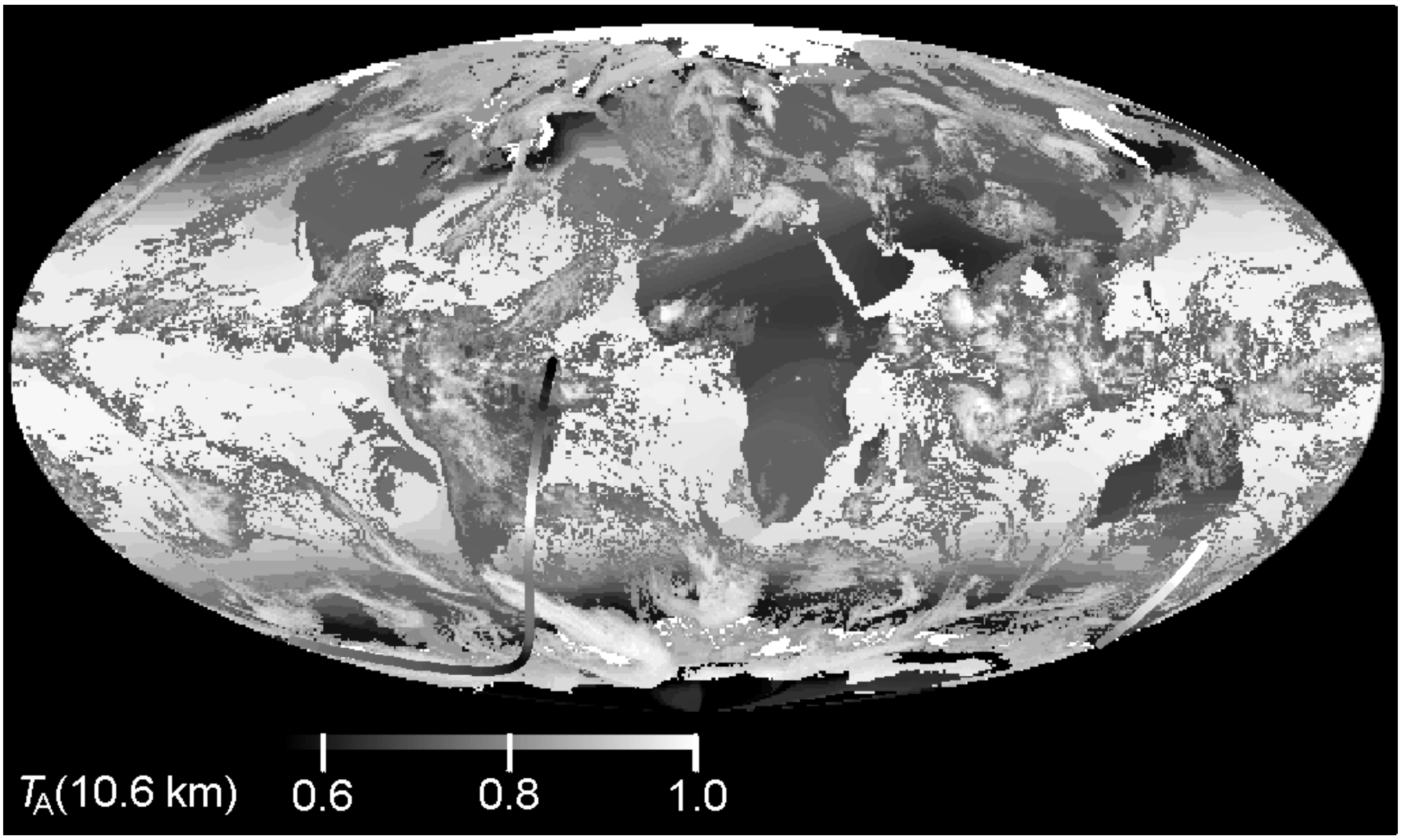

Figure 10. Correlation between aerosol transparency for the radiation path with tangent altitude 10.6 km and cloud distribution along the limb. The map is taken from [9] for May, 05, 2004, $06^h$ UT.

At the eclipse of October, 28, aerosol extinction rapidly increases in the lower troposphere. The picture being observed (except the near-equatorial area) can be described by the simplest stair-like model:

$$\begin{aligned} T(h) &= 0, \quad h \le h_A, \\ T(h) &= 1, \quad h > h_A, \end{aligned} \qquad (27)$$

where the upper border altitude of the aerosol extinction layer $h_A$ is equal to 5.0 km. At these latitudes the aerosol extinction coefficient exceeds $10^{-2}$ km$^{-1}$ and atmosphere is not transparent for tangent rays expanding there. It is clearly seen that aerosol extinction coefficient in troposphere decreases with altitude many times faster than molecular one.

**5. Discussion and conclusion**

Atmospheric aerosol extinction measurement was the primary goal of the total lunar eclipses photometry. The analysis led us to divide the Earth's atmosphere on three layers with different optical properties during the lunar eclipses:

1) Lower troposphere, from the ground till the altitudes about 5-6 km: the aerosol extinction of the tangent radiation is strong, the measured data is the average over large areas along the limb. This layer is the principal for the brightness of deep umbra regions. Being observed from the Moon, the Sun is situated behind the central regions of the Earth disk and its emission is refracted in a wide zone above the Earth's limb. We can measure the general optical properties of this layer, first of all, its increase from the poles to the equator, which is seen for both eclipses. At altitudes lower than 3 km the transparency is near to zero.

2) Upper troposphere, from 5-6 to 10-12 km: the aerosol extinction appears only in definite areas above the limb, the correlation with cyclones is good for polar and mid-latitude locations and bad near the equator. It seems to be most interesting layer for the lunar-eclipse scanning.
3) Tropopause and lower stratosphere, above 10-12 km: the measured umbra darkening factor is in good agreement with theoretical model data for the molecular atmosphere without aerosol extinction. Above 20 km the refraction is very small and radiation transferring at these altitudes is undetectable on the Moon due to large sky background variability near the umbra border.

The upper value of troposphere aerosol layer, after which the atmosphere becomes transparent for the tangent solar radiation, is turned out to vary from 5 km (clear atmosphere case at October, 28) to 10 km and higher over the polar cyclones at May, 04. These values are in good agreement with the estimations of aerosol tangent transparency based on twilight sky photometry [10], providing the altitude values from 7 to 12 km for central Russia. Polarization measurements of scattered radiation during the twilight [11] detected the admixture of aerosol scattering at the twilight layer altitude about 15 km. But if we take into account the effective thickness of this layer (about the uniform atmosphere layer scale, 6-8 km), the results will be in good agreement again.

The observations were conducted before the epoch of solar activity minimum. A number of observations through XIX and XX centuries (see [4] for review) had shown that during the solar activity minimum period the brightness of lunar eclipses rapidly changes (decreases). Having compared the atmospheric aerosol data obtained here with the same data after the solar activity minimum (obtained during future eclipses) we can investigate the correlation of atmospheric aerosol properties at different altitudes with solar activity.

However, the layer distribution described above can depend on the spectral band. If we choose shorter wavelength, the atmosphere transparency will decrease and we will be able to detect weaker aerosol clusters, but the contribution of scattered radiation will increase, that will bring additional problems. The picture can rapidly change if we measure the Moon brightness in a narrow band inside the absorption line of atmosphere gas. Comparison of this brightness with the one measured in a nearby spectral band will bring the information about the distribution of this gas in the atmosphere, both horizontal (along the limb) and vertical, adding the space experiments data. The spectroscopy of lunar eclipse can be also useful for atmospheric research.

**Acknowledgements**

The authors would like to thank V.I. Shenavrin and V.F. Esipov (Sternberg Astronomical Institute, Moscow, Russia) for their help during the observations, N.N. Shakhvorostova (Astro-Space Center of Lebedev’s Physical Institute, Moscow, Russia) and Yu.I. Velikodsky (Kharkov Astronomical Observatory, Ukraine) for the useful remarks.